\documentclass{PoS}
\usepackage[many]{tcolorbox}
\usepackage{xspace}
\usepackage{listings}
\newcommand{\modelname}[1]{\texttt{#1}\@\xspace}
\newcommand{\sarah}{\texttt{SARAH}\@\xspace}

\newcommand{\fs}{\texttt{FlexibleSUSY}\@\xspace}

\newcommand{\HSSUSY}{\modelname{HSSUSY}}
\newcommand{\SplitMSSM}{\modelname{SplitMSSM}}
\newcommand{\thdmf}{\modelname{THDMIIMSSMBCFull}}
\newcommand{\hthdm}{\modelname{HTHDMIIMSSMBC}}
\newcommand{\hgthdm}{\modelname{HGTHDMIIMSSMBCFull}}

\newcommand{\susyhd}{\texttt{SusyHD}\@\xspace}

\newcommand{\softsusy}{\texttt{SOFTSUSY}\@\xspace}
\newcommand{\micromegas}{\texttt{micrOMEGAs}\@\xspace}

\newcommand{\Himalaya}{\texttt{Himalaya}\@\xspace}

\newcommand{\multinest}{\texttt{MultiNest}\@\xspace}

\newcommand{\gambit}{\texttt{GAMBIT}\@\xspace}

\newcommand{\fstwo}{\fs 2.0\@\xspace}

\newcommand{\famu}{\texttt{FlexibleAMU}\@\xspace}

\newcommand{\feft}{\texttt{Flex\-ib\-le\-EFT\-Higgs}\@\xspace}

\newcommand{\code}[1]{\lstinline|#1|}  

\newcommand{\ptitle}[1]{\emph{#1}}
\renewcommand{\ptitle}[1]{}

\newcommand{\mhalf}{\ensuremath{M_{1/2}}\xspace}
\newcommand{\mzero}{\ensuremath{m_0}\xspace}

\title{\fs: Precise automated calculations in any BSM theory}

\ShortTitle{FlexibleSUSY}

\author{\speaker{Peter Athron} \\ ARC Centre of Excellence for
  Particle Physics at the Terascale, School of Physics, Monash
  University, Melbourne, Victoria 3800, Australia \\ E-mail:
  \email{peter.athron@coepp.org.au}}

\author{Markus Bach
}

\author{Dylan Harries
}

\author{Wojciech Kotlarski
}

\author{Thomas Kwasnitza
}

\author{Jae-hyeon Park 
}
 \author{Tom Steudtner
 }

\author{Dominik St\"ockinger
}

\author{Alexander Voigt
}

\author{Jobst Ziebell
}

\abstract{\fs is a software package for various calculations in any
  model of physics beyond the standard model (not just any
  supersymmetric model).  \fs can solve boundary value problems and
  uses this to find $\overline{DR}/\overline{MS}$ parameters and calculate the Higgs
  and BSM particle masses, as well as other observables. \fs is designed to
  be adaptable, fast, precise and reliable. We describe \fs with
  particular emphasis on recent developments and the state of the art
  Higgs mass calculations it can perform. We also show some
  applications to illustrate how it can be used to obtain interesting
  physics results with the highest precision possible and with
  remarkable speed.}

\FullConference{39th International Conference on High Energy Physics\\
		4-11 July 2018\\
		Seoul, South Korea}

\begin{document}

\section{Introduction}
\noindent A major difficulty in investigating the phenomenology of a
novel extension of the standard model is that for meaningful predictions
of observables one must perform many dedicated calculations, often at
the 1- or 2-loop level and then implement these calculations in a
programming language so that the parameter space can be explored. Therefore
automation of these steps can be extremely useful for
phenomenology. \fs
\cite{Athron:2014yba,Athron:2016fuq,Athron:2017fvs} is an easy-to-use
software package which generates a C++ spectrum generator and C++
calculations of observables, for any BSM model.  Here we will describe
\fs and many of its uses.
\section{\fs}
 \noindent \fs uses \sarah \cite{Staub:2009bi} to obtain
 model-specific expressions and corrections.  These analytic
 expressions obtained in Mathematica, along with additional
 calculations and corrections from \fs and other sources (discussed
 later), are then translated to C++ and implemented in a code for
 spectrum generator\footnote{This spectrum generator contains some
   numerical routines from \softsusy \cite{Allanach:2001kg} and is
   unit tested against code-pieces from the MSSM and NMSSM versions.}
 and observable calculations which is fast, adaptable and
 reliable. The form of the spectrum generator and what calculations
 are included are specified by the \fs model file.  \fs is distributed
 with a large variety of model files already written for both SUSY
 models (e.g. MSSM, NMSSM, UMSSM, E6SSM and MRSSM) and non-SUSY models
 (e.g. THDM, MDM, BLSM, SplitMSSM).  The user can immediately use
 these model files to create a spectrum generator for that model
 with three simple commands, e.g.\
\begin{lstlisting}[language=bash]
$ ./createmodel --name=MRSSM
$ ./configure --with-models=MRSSM
$ make
\end{lstlisting}
The \fs spectrum generator may then be run with the following:
\begin{lstlisting}[language=bash]
$ cd models/MRSSM
$  ./run_MRSSM.x --slha-input-file=LesHouches.in.MRSSM
\end{lstlisting}
\fs also provides many prebuilt spectrum generators
\cite{prebuild-specgen} and even a web-interface for generating mass
spectra for selected models without ever downloading \fs
\cite{web-interface}.

In the MSSM and NMSSM \fs can output SLHA \cite{Skands:2003cj} and
SLHA2 \cite{Allanach:2008qq} files for interfacing with other codes.
In other models a similar SLHA-like interface can be used if the other
codes have also been setup to work with that model by using \sarah,
with the same \sarah model files as \fs.  For example \fs linked to \micromegas
\cite{Belanger:2001fz} can be used to investigate dark matter \cite{Athron:2015vxg}, with
the parameter space sampled efficiently by also interfacing to
\multinest \cite{Feroz:2008xx}.  \fs can also be used in the global
fitting tool \gambit \cite{Athron:2017ard} where it is interfaced with
a huge array of observable calculators and samplers.

2-loop RGEs, and full 1-loop self-energy and tadpole corrections are
obtained using model-specific analytical expressions that are
generated using \sarah. In addition \fs also implements its own
corrections that are also applied to all models.  These include a
1-loop calculation of threshold corrections for extracting
$\alpha_s(m_Z)$ and $\alpha_e(m_Z)$ in the considered extension of the
SM, a new (partial 2-loop) calculation of the weak mixing angle $\sin
\theta_W$ which allows us to obtain the $SU(2)_W$ and $U(1)_Y$ gauge
couplings, $g$ and $g^\prime$ respectively, and pure QCD 2-loop
corrections to the running top and bottom Yukawa couplings. With these
precision corrections \fs provides state-of-the-art predictions of the
mass spectra in any novel extension of the SM chosen by the
user.

However in addition to the calculations applicable to any model, \fs
contains many model specific corrections where there have been recent
improvements.  Therefore \fs includes: 3-loop RGEs in the SM and MSSM;
2-loop (SUSY) QCD corrections to top, bottom Yukawas in SM and MSSM
\cite{Allanach:2014nba}; 2-loop (SUSY) QCD corrections to strong gauge
coupling in SM, MSSM \cite{Allanach:2014nba}; 2-loop fixed order Higgs
mass corrections SM, MSSM, NMSSM; 3-loop fixed order Higgs mass
corrections SM, MSSM (via \Himalaya \cite{Harlander:2017kuc}); and
3-loop Higgs mass corrections in Split-MSSM \cite{Benakli:2013msa}.
Thus \fs also provides state-of-the-art calculation in well-studied
models like the MSSM, NMSSM.

\begin{figure}[tbh]
  \centering \includegraphics[width=0.35\textwidth]{%
    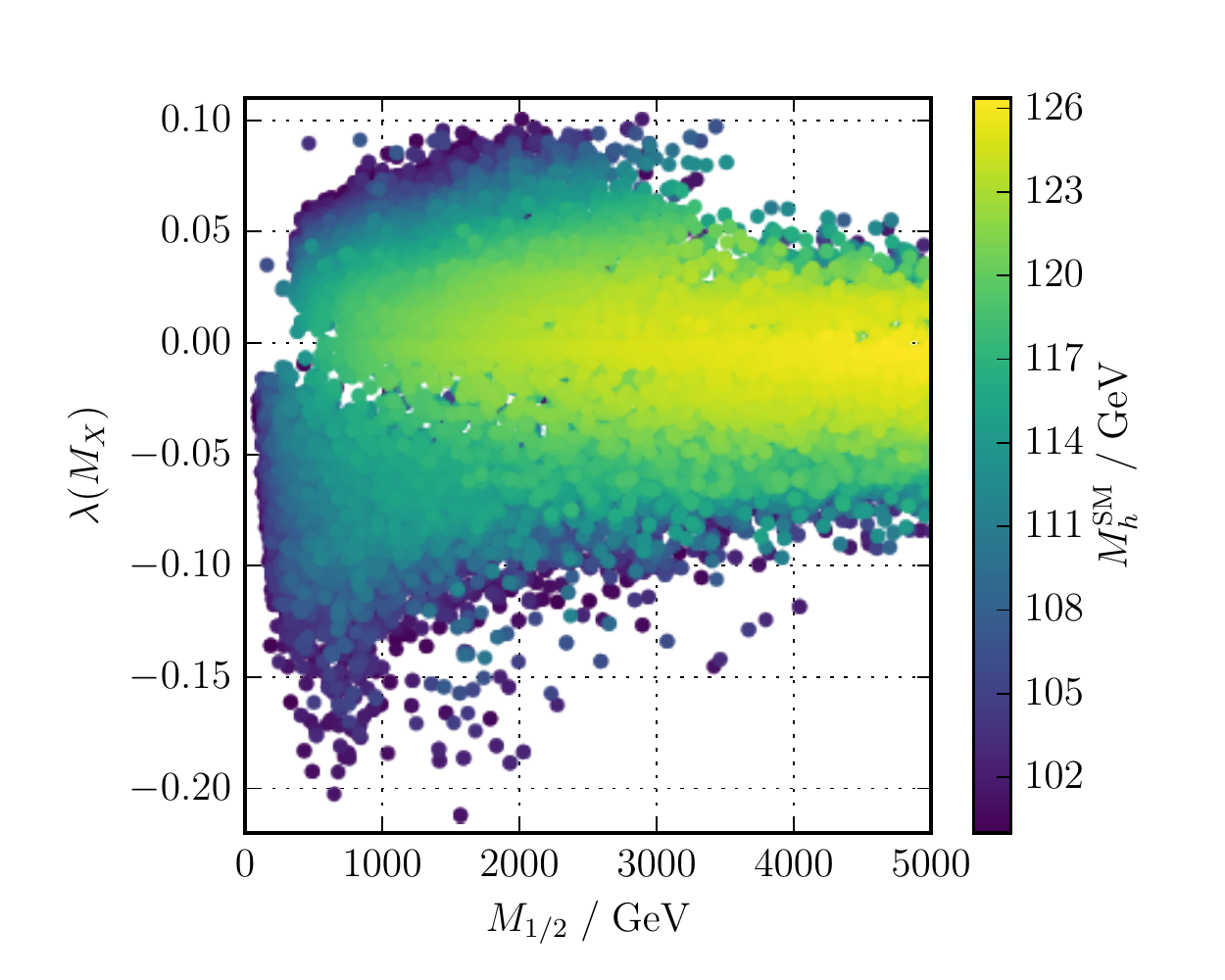}
  \includegraphics[width=0.35\textwidth]{%
    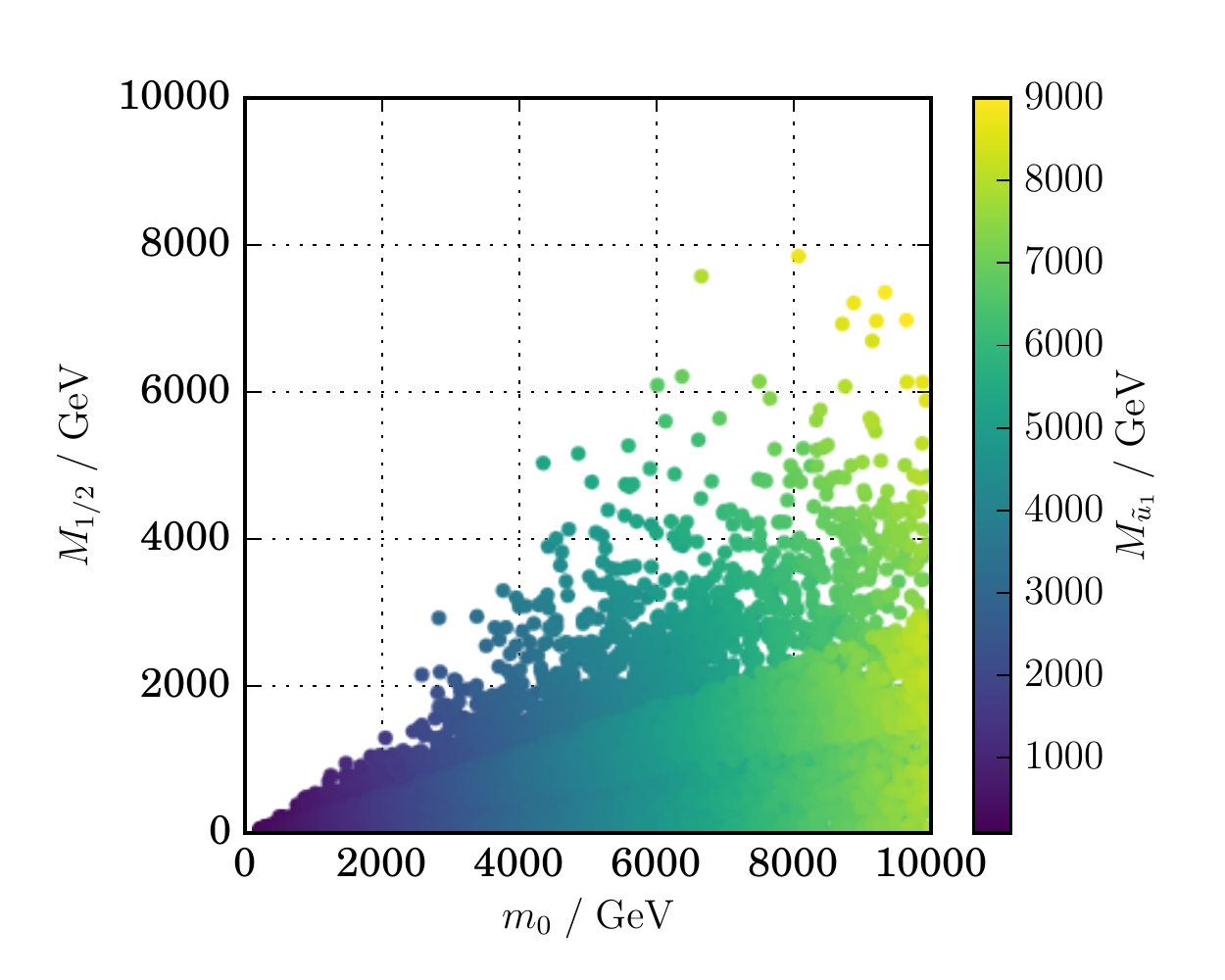}
   \caption{CNMSSM (left) and CE$_6$SSM (right) solutions obtained using
    \fstwo and \multinest 3.1 shown in the $\mzero-\mhalf$ plane with
    the SM-like Higgs mass and the lightest up-squark mass as a color
    contour respectively. These solutions are found using the
    semi-analytic solver for the BVP.}
   \label{fig:constrained_solutions}
\end{figure}

The \fs spectrum generators are also highly adaptable. 
Boundary value problems (BVPs) with boundary conditions at both the
high scale and the low scale can be solved with the two-scale fixed
point iteration, used widely in spectrum generators for the MSSM and
NMSSM.  However in \fs it can now be replaced with the semi-analytic
solver, which makes use of semi-analytic solutions to the RGEs so that
the electroweak symmetry breaking conditions may be re-written in
terms of (possibly universal) parameters defined at the high scale.
This makes it possible to find solutions in, for example, fully
constrained versions of the NMSSM and E$_6$SSM (see
Fig.\ref{fig:constrained_solutions}), without scanning over additional
parameters.  This is not possible in any other public spectrum
generator.  Both the boundary conditions and the actual definition of
the boundary scale may be chosen by the user in the model file, where
they may write in many different types of functions of the parameters
(including e.g.\ trigonometric functions, dilogarithms and the
Kronecker-delta).  For example the high scale can be required to be
the scale where the electroweak gauge couplings are equal ($g_1 =
g_2$), or where the bottom and tau Yukawa couplings are unified
($y_\tau = y_b$) or defined by some more complicated expression. Multiple
models can also be used to build a tower of effective field theories at the
C++ level (full automation is under development).

Since the Higgs boson mass has been measured with an uncertainty is
that is already significantly smaller than the theoretical uncertainty, \fs
has a number of very sophisticated Higgs mass calculations. For all
models \fs implements the \feft \cite{Athron:2016fuq} algorithm, which
resums logarithms at NLL precision ensuring precise predictions when
the new physics scale is high, while matching the 1-loop
precision\footnote{A 2-loop fixed order calculation for any model,
  using a \sarah extension \cite{Goodsell:2015ira} is
  currently under development.} of the fixed-order calculations when
the new physics appears at the electroweak scale\footnote{A version of
  this procedure with 2-loop matching to provide NNLL resummation is
  currently under development.}.  We therefore recommend users employ
this approach in all cases unless there is a more precise
model-specific EFT or fixed-order calculation (depending on the new
physics scale) available.  For specific models we have additional
fixed order corrections mentioned above, allowing an improved
prediction of the Higgs mass, for example at the 3-loop level in the
MSSM via \Himalaya \cite{Harlander:2017kuc}.  We also provide special
model files where matching conditions between the specific high scale
and low scale model are implemented.  The \fs spectrum generators
created from these specific model files then perform precise Higgs
mass calculations with resummed logs. Currently \fs has:
\begin{itemize}
\item \HSSUSY \cite{Bagnaschi:2015pwa}: MSSM calculation assuming a SM
  EFT, predicts $M_h$ at full NLO with full NLL ressummation + NNLO with NNLL ressumation at ${\cal O}(\alpha_t(\alpha_s+\alpha_t) + (\alpha_t+\alpha_b)^2 +\alpha_\tau\alpha_b + \alpha_\tau^2)$
  + (very recent) N$^{3}$LO corrections
  with N$^{3}$LL at ${\cal O}(\alpha_t\alpha_s^2)$ resummation\footnote{With these recent
    corrections \cite{Harlander:2018yhj} this C++ Higgs mass calculator is the most precise EFT
    calculation of the Higgs mass available for the MSSM, with a higher
    precision than the next-best competitor \susyhd
    \cite{Vega:2015fna}.}.
  \item \SplitMSSM: MSSM calculation assuming a SplitSUSY EFT, predicts $M_h$ at full NLO + NNLO ${\cal O}(\alpha_t\alpha_s)$ with full NLL resummation. 
  \item \thdmf: MSSM calculation assuming a THDMII EFT, predicts  $M_h$ at full NLO (+ partial NNLO ${\cal O}(\alpha_t\alpha_s)$) with full NLL resummation. 
   \item \hthdm: MSSM calculation assuming an EFT with THDMII + light Higgsinos states, predicts $M_h$ at full NLO (+ partial NNLO ${\cal O}(\alpha_t\alpha_s)$) with full NLL resummation. 
\item \hgthdm: MSSM calculation assuming an EFT with THDMII + light Higgsino + light gaugino states, predicts $M_h$ at full NLO (+ partial NNLO ${\cal O}(\alpha_t\alpha_s)$) with full NLL resummation. 
\end{itemize}
As well as the spectrum and Higgs mass calculations, \fs has also
implemented additional observable calculations. \fs includes a \famu
calculation of the anomalous magnetic moment of the muon at 1-loop,
including fermions and scalars in the loops.  Additional 2-loop
contributions and vector contributions will be included in a future
update.  \fs also includes a 1-loop calculation of lepton EDMs which
can provide important constraints when the parameters include complex
phases.  The new extraction of $\sin\theta_W$ also allows us to
predict the mass of the $W$ boson, instead of using it as an input to
obtain $\sin\theta_W$.  \fs can also provide the effective vertices
for $h \rightarrow\gamma \gamma$ and $h \rightarrow gg$ decays (where $h$ can be the SM
Higgs or any neutral scalar in the SM extension). A
full loop-level calculation of Higgs decays and decays of BSM states
is under development, with a release anticipated soon.

\section{Conclusions}
\noindent \fs is a software package for calculating mass spectra and a growing
set of observables in a fast, flexible, precise and robust manner. \fs
is easy to use and can be readily interfaced with other codes to do
detailed phenomenological investigations, and even global fits in both
novel and well-known extensions of the standard model.

\end{document}